\documentclass[aps,prev,twocolumn,superscriptaddress,preprintnumbers,nofootinbib]{revtex4-1}
\pdfoutput=1

\usepackage{graphicx}
\usepackage{bm}
\usepackage{times}
\usepackage{slashed}
\usepackage{amssymb}
\usepackage{color}
\usepackage{slashed}
\usepackage{lipsum}
\usepackage{subfigure}
\usepackage{multirow}
\usepackage{amsmath}
\usepackage{array}
\usepackage{varwidth}
\usepackage{float}

\def\beq{\begin{equation}}
\def\eeq{\end{equation}}
\def\bea{\begin{eqnarray}}
\def\eea{\end{eqnarray}}

\def\eqref#1{Eq.~(\ref{eq:#1})}

\newcommand*{\figref}[1]{Fig.~\ref{fig:#1}}
\newcommand*{\figlab}[1]{\label{fig:#1}}

\begin{document}

\title{Coherent transition radiation from the geomagnetically-induced current in cosmic-ray air showers: Implications for the anomalous events observed by ANITA}

\author{Krijn D. de Vries}
\email{krijn.de.vries@vub.ac.be}
\affiliation{Vrije  Universiteit  Brussel,  Dienst  ELEM,  IIHE,  Pleinlaan  2,  1050,  Brussel,  Belgium}

\author{Steven Prohira}%
 \email{prohira.1@osu.edu}
 \affiliation{Center for Cosmology and AstroParticle Physics (CCAPP), The Ohio State University, Columbus OH, 43210}

\begin{abstract}
We show that coherent transition radiation from the electrically-neutral transverse geomagnetic current (CTR-GM) in a cosmic-ray air shower provides a natural, standard model, explanation to the recent ``anomalous'' events observed by the ANITA detector. We demonstrate that for zenith angles less than $\sim$70 degrees, combined with high surface elevation, the inclusion of CTR-GM can significantly alter the emitted electric field from a cosmic-ray air shower. CTR-GM therefore has to be included in radio emission models to provide a full description of the radio emission from a high-energy cosmic-ray air shower traversing a dielectric boundary.
\end{abstract}

\maketitle

{\bf Introduction---}
During propagation through the atmosphere, the electrons and positrons of a cosmic ray air shower will be deflected in Earth's magnetic field, inducing a net transverse current in the shower front. This results in geomagnetically induced radio emission up to GHz frequencies~\cite{Kahn,Olaf,cher1,cher2}, which allows the shower to be detected with radio instruments. A second radio emission source is the net negative excess charge in the shower, first predicted by Askaryan in 1962~\cite{Askaryan} and experimentally confirmed in 2001~\cite{Gorham}.

Detection of high energy cosmic rays using radio is currently a well established method~\cite{radioreview1,radioreview2}, with development stretching back to the initial efforts of the 1960's~\cite{allen_rev}. In addition to cosmic-ray air shower radio detectors, several radio detectors are currently under development to probe particle cascades induced by high-energy ($>$~10~PeV) cosmic neutrinos interacting in more dense media such as ice or rock~\cite{anita_instrument,ARA,ARIANNA,GRAND}. 

In recent works~\cite{CosmicRayRadio,Motloch,Motloch2} another important emission mechanism is discussed, coherent transition radiation (CTR) from a high-energy particle cascade traversing different media. The considered source of the transition radiation is the net excess charge predicted by Askaryan (CTR-A). Other works considering particle cascades moving through dielectrics concern the EXTASIS experiment~\cite{extasis}, for which the emission at ground based cosmic-ray detection set-ups due to the absorption of the cosmic-ray air shower by Earth was treated in Ref.~\cite{extasis_th}. Recently, CTR-A was confirmed experimentally at the Telescope Array Electron Light Source facility, where the emission from a high-energy electron beam leaving the accelerator was quantified in detail~\cite{SA-TA-ELS}, earlier works on CTR-A using electron beams are found in Refs.~\cite{Argonne,SLACsalt}.

In this work, we present a second CTR contribution that is expected for cosmic-ray air showers hitting a boundary surface. We show that strong coherent transition radiation from the geomagnetically-induced current (CTR-GM) can be expected once a significantly large particle number crosses a boundary surface. We also show that this condition in general is satisfied for high-energy cosmic-ray air showers with shallow zenith angles ($\lesssim70^\circ$) incident upon high surface elevations ($\gtrsim$2.5~km). If these criteria are not satisfied, the particle content at the boundary is too small for the coherent transition radiation to be significant~\cite{CosmicRayRadio,extasis_th}. 

We discuss the obtained results in the context of the two so-called `anomalous events'  detected by the stratospheric balloon-borne ANITA instrument~\cite{unusual1}\cite{unusual2}. Though primarily a neutrino detector, ANITA also observes the emission from downward-going cosmic-ray air showers after reflection from the ice. The anomalous cosmic-ray like events detected by ANITA have the same polarization as a typical cosmic-ray, but an inverted polarity. {\it Polarization} is a measure of the plane of oscillation of the electric field,  while {\it polarity} is a measure of the sign of the dominant peak(s) of this field. Consequently, the inverted polarity of the anomalous events has been interpreted as emission from an upward-going shower of some kind. 

Such an upward moving cascade, however, is only possible for primary particles having traversed a long path through the earth. Plausible standard-model explanations for such events are largely ruled out by ANITA exposure limits~\cite{exposure}, and other explanations require physics beyond the standard model~\cite{bsm1,bsm2,bsm3,bsm8,bsm4,bsm6,bsm7,bsm9,bsm10,bsm11}.

In this work, we show that CTR-GM from a down-going cosmic ray shower provides a natural explanation for the observed inverted polarity signals. We show that CTR-GM significantly affects the expected electric field pulse shapes and that the two anomalous events have cascade geometries for which strong coherent transition radiation is expected, which is not the case for the majority of the cosmic-ray events observed by the ANITA detector.
\\\\

{\bf Coherent transition radiation---} To calculate the coherent transition radiation from the geomagnetically-induced air shower current, we follow the approach presented in Ref.~\cite{CosmicRayRadio}. In what follows, we denote $z$ as the axis normal to the ice surface, the subscript $b$ refers to the air/ice boundary, and primed quantities are `retarded' or `emission' times.
\begin{figure}[tb]
  \centering

    \includegraphics[width=0.48\textwidth]{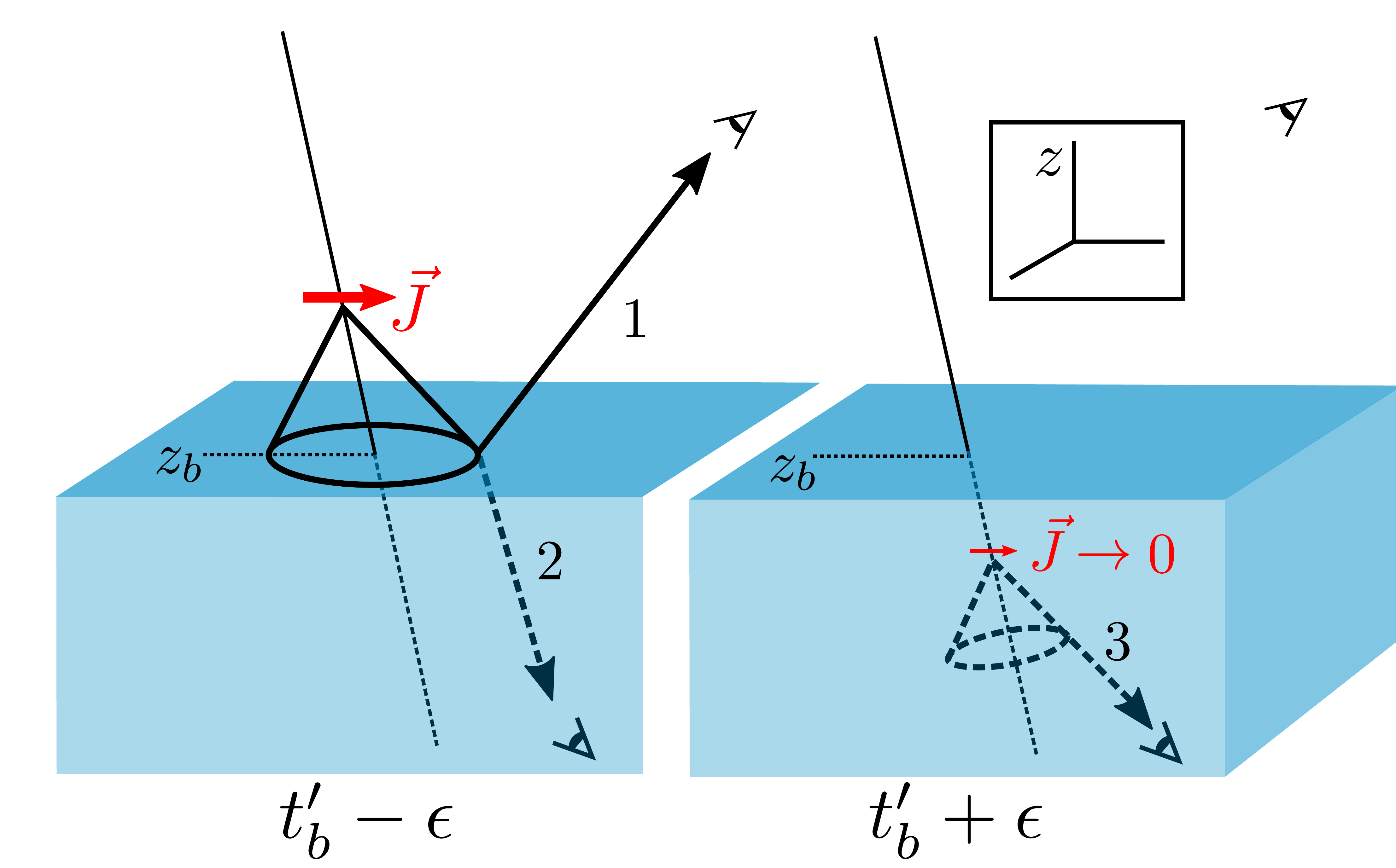}
  \caption{The geometry for coherent transition radiation just before (left) and just after (right) the relativistic shower front crosses the air/ice boundary $z_b$ at retarded time $t^\prime_b$. The geomagnetic current is indicated by $\vec{J}$. The direct emission from this current is given by the solid black cone on the left figure, which reflects up (label 1) for an in-air observer and refracts down for an in-ice observer (label 2). At time $t^\prime_b+\epsilon$, the path to the in-air observer has vanished.}
  
  \figlab{geometry3}
\end{figure}

Since the particle cascade is moving relativistically, the emission will be boosted along its direction of motion. In the following we will therefore only consider emission in the forward direction. In the left half of \figref{geometry3}, the forward emission from a cosmic-ray air shower while propagating in air is illustrated by the full black cone. The signal gets `split' at the boundary $z_b$, where part of the emission gets transmitted into the ice given by the dashed line (label 2) and part of the signal gets reflected off the surface, shown by the full line (label 1). After the cascade penetrates the ice at time $t_b$ (\figref{geometry3}, right), the reflected path completely vanishes and only the direct path remains (label 3). 

From~\figref{geometry3}, it immediately follows that the observed emission depends strongly on the observer location. For an in-ice observer, one has to consider paths 2 and 3, where for an in-air observer only path 1 has to be considered. In the following, we derive the expected fields due to coherent transition radiation for both situations separately.

{\bf CTR for an in-ice observer---}
The in-air potential obtained at an infinitesimal distance $\epsilon$ above the boundary is observed through a completely different refracted path (\figref{geometry3}, left, label 2), compared to the emission emitted an infinitesimal distance $\epsilon$ below the boundary (\figref{geometry3}, right, label 3). A direct consequence of this discontinuity in the path length is a discontinuity in the observed potentials, leading to strong emission from the boundary, coherent transition radiation. It also follows that this is a geometrical effect, and as such is not limited to the net excess charge predicted by Askaryan, but applies equally well to the geomagnetically-induced air shower current.

To calculate the transition radiation from the geomagnetically-induced current for a typical cosmic-ray air shower, we consider the three dimensional current distribution given by $J^x(t',\vec{r},h)=N_e(t') e v_d w(\vec{r},h)$. Here, $N_e(t')$ denotes the number of leptons at the emission time $t'$, to be observed at the observer time $t$. The drift velocity $v_d=0.04~c$ is the average velocity induced by Earth's magnetic field for electrons and positrons in a typical cosmic-ray air shower~\cite{Wer12}. The function $w(\vec{r},h)$ gives the particle distribution within the charge cloud, where $\vec{r}$ denotes the lateral distance and $h$ the longitudinal distance within the cloud which moves by definition with the speed of light along the cascade axis. The total particle number, as well as the particle distributions within the charge cloud are parameterized following Ref.~\cite{Wer12} and the detailed parameterizations are presented in the supplementary materials.

Closely following the formalism for a net excess charge presented in Ref.~\cite{CosmicRayRadio}, the potential for the geomagnetically-induced air shower current crossing a boundary surface is given by,
\bea
A^{x}(\vec{x},t)&&=\frac{\mu_0}{4\pi}\left.\int \mathrm{d^2}\vec{r}\;\mathrm{d}h\;{\cal T}_{air-ice} \frac{J^{x}(t',r,h)}{ |{\cal D}_{2}|}\theta(z-z_b)\right|_{t'}\nonumber\\
&&+\frac{\mu_0}{4\pi}\left.\int \mathrm{d^2}\vec{r}\;\mathrm{d}h\; \frac{J^{x}(t',r,h)}{ |{\cal D}_{3}|}\theta(z_b-z)\right|_{t'}.
\eea
Here, ${\cal T}_{air-ice}$ denotes the Fresnel transmission coefficient, and the different path lengths are included in the retarded distance ${\cal D}=L\frac{dt}{dt'}$, where $L$ denotes the optical path length from the emission point at emission time $t'$ to the observer, where the signal arrives at the observer time $t$. The electric fields are now obtained through the standard relation $\vec{E}=-dA^{0}/d\vec{x}-d\vec{A}/d(ct)$. These derivatives work on all terms under the integral. The specific contribution where the derivatives operate on the Heaviside step function $\theta(z)$ is called CTR and gives rise to the fields

\bea
{E}^{x}(\vec{x},t)&&=\frac{\mu_0}{4\pi}\lim_{\epsilon \rightarrow 0} \int \mathrm{d^2}\vec{r} \left.{\cal T}_{air-ice}\frac{ J^{x}(t',r,h)}{ |{\cal D}_{2}|}\right|_{z=z_b+\epsilon}\nonumber\\
&&-\frac{\mu_0}{4\pi}\lim_{\epsilon \rightarrow 0} \int \mathrm{d^2}\vec{r}\left.\frac{J^{x}(t',r,h)}{ |{\cal D}_{3}|}\right|_{z=z_b-\epsilon}.
\eea

{\bf CTR for an in-air observer.---}
The derivation for coherent transition radiation for an in-air observer is similar to the derivation for an in-ice observer outlined above. For an in-air observer, however, the reflected path through which the potential is observed (\figref{geometry3}, left, label 1) instantly vanishes below the boundary. Though the (backward) emission from the cascade propagation below the ice reaches the detector, due to relativistic beaming this component is negligible compared to the reflected component from the in-air emission. As such, it is safe to ignore this in-ice contribution to the potential. We continue to use the term ``transition radiation,'' however, because the transition from air to ice is explicitly responsible for the induced shock in the potential. The potential from the geomagnetically-induced air shower current for an in-air observer is thus given by,
\beq
\left.A^{x}(\vec{x},t)=\frac{\mu_0}{4\pi}\int \mathrm{d^2}\vec{r}\;\mathrm{d}h\; {\cal R}_{air-ice}\frac{J^{x}(t',r,h)}{ |{\cal D}_{1}| }\theta(z-z_b)\right|_{t'},
\eeq
using the Fresnel reflection coefficient ${\cal R}_{air-ice}$. We subsequently obtain the field due to the vanishing of the potential at the boundary,
\beq
\vec{E}^{x}(\vec{x},t)=\frac{\mu_0}{4\pi}\lim_{\epsilon \rightarrow 0} \int \mathrm{d^2}\vec{r} \left.{\cal R}_{air-ice}\frac{J^{x}(t',r,h)}{ |{\cal D}_{1}|}\right|_{z=z_b+\epsilon}.
\eeq

\begin{figure}[t]
  \centering
  \includegraphics[width=0.48\textwidth]{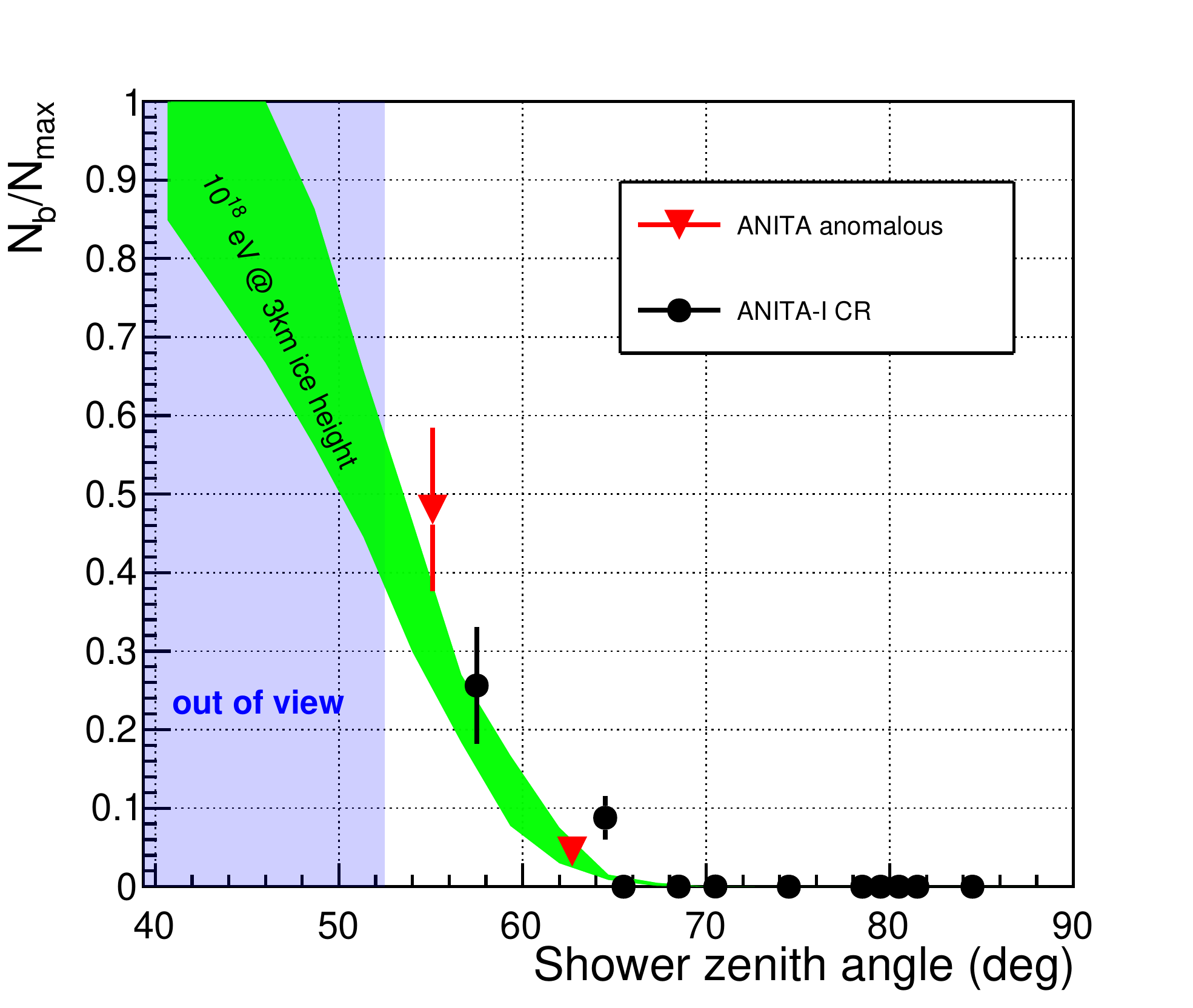}
  
  \caption{The ratio of the number of charged particles hitting the ice boundary ($N_b$) to the maximum number of charged particles in the shower ($N_{max}$), versus zenith angle of the shower. Both the ANITA-I cosmic rays as well as the ANITA-I and ANITA-III anomalous events are shown. The error on the ANITA-1 CR sample is due to the reported error in the energy estimate. The shaded region is excluded by ANITA's field of view.}
  \figlab{anita_sensitivity}
\end{figure}

{\bf The ANITA anomalous events.---} In this section we investigate if coherent transition radiation from the geomagnetically-induced air shower current can provide an explanation for the ANITA anomalous events. For strong coherent transition radiation to occur, a significant part of the cosmic-ray air shower has to hit Earth's surface. From Fig.~7 in Ref.~\cite{CosmicRayRadio}, it follows that this restricts us to cosmic-ray air showers with zenith angles below $\sim 70$~degrees for the typical environments in which the ANITA detector operates. 

To quantify this, a set of Monte-Carlo (MC) showers has been made using the CX-MC-GEO package presented in Ref~\cite{Wer12}. This package allows to obtain the three dimensional charge and current distributions located inside the high-energy cascade front and is based on the full MC cascade mode of CONEX~\cite{Conex1,Conex2}. The showers have been produced in the zenith angle range from 40-80~degrees, containing 10 simulations for each degree interval. The considered showers are induced by a $10^{18}$~eV proton primary, and the air-ice boundary is chosen at 3~km above sea level similar to conditions at Antarctica. 

The results are shown by the green plane in~\figref{anita_sensitivity}. Here we plot the number of particles hitting the air-ice boundary, $N_b$, with respect to the maximum number of shower particles, $N_{max}$. From this we indeed confirm that for zenith angles larger than 70~degrees the particle content at the boundary becomes small. 

Additionally, the shaded area in~\figref{anita_sensitivity} shows the range of zenith angles excluded by the ANITA antenna array field of view~\cite{anita_instrument}. To investigate in more detail if the ANITA anomalous events lie within the region of interest for coherent transition radiation to be significant, we ran 10 Monte-Carlo showers for each anomalous event within their given reconstruction errors. These errors are obtained by using the event parameters published by the ANITA collaboration~\cite{unusual2}, including energy (the reported energies for the anomalous events, provided for a direct shower, have been scaled by the empirically determined surface reflectivity coefficients attained on recent ANITA flights~\cite{a4_hc}), zenith angle, and surface elevation.  A similar procedure has been made for the normal cosmic-ray air shower events detected during the ANITA-1 flight~\cite{harm,anita_private}.

From~\figref{anita_sensitivity}, it indeed follows that the ANITA anomalous events lie within the region of interest, separated from the majority of the ANITA cosmic-ray sample. For the CTR-GM signal to be significant, a combination of three variables of each event has to be considered: the event must be high energy, arrive at a zenith angle~$\lesssim$70 degrees, and impact the surface at a high elevation. It happens that in these variables, the ANITA anomalous events are within the region of interest and live in the tails of the typical CR distribution.

To quantify if coherent transition radiation can be an explanation for the ANITA anomalous events, as well as the normal cosmic-ray air showers observed in this region, in~\figref{results}, we show the expected field for one of the ANITA anomalous events with a reconstructed zenith angle of 55~degrees and a surface elevation of 2.7~km, that is seen by an observer located 30~km above the air-ice boundary. The field is simulated at different viewing angles of $\alpha=0,1.7,3.9$~degrees with respect to the specular angle of the cascade axis. It follows that, depending on the observer geometry, the expected transition radiation can be large compared to the geomagnetic emission in air, inverting the field polarity (\figref{results} (b)). 
Furthermore, the time-ordering of the peaks can be reversed, (\figref{results} (a)). These effects are absent for observer geometries further out from the specular angle (\figref{results} (c)). 
For comparison, in~\figref{results} (d), we show the expected emission observed at a viewing angle of $\alpha=1.7$~degree for a primary of the same energy and surface elevation, only adapting the zenith angle to 70 degrees, typical of an ANITA-1 CR. As expected in this situation, the transition radiation becomes negligible.

\begin{figure}[t]
  \centering
  \includegraphics[width=0.51\textwidth]{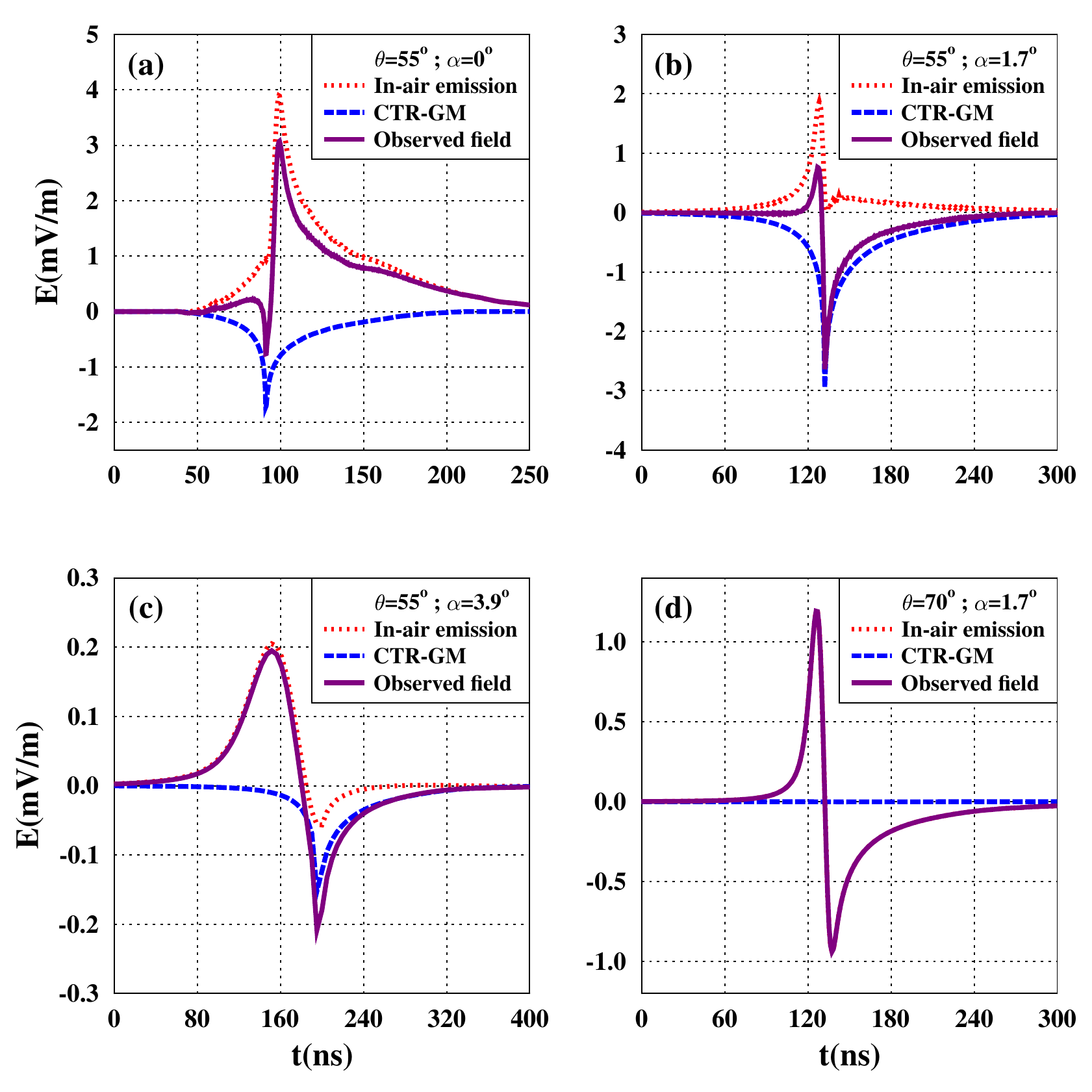}
    \caption{The expected radio emission (full purple lines) from a down-going cosmic ray air shower hitting an air-ice boundary at 3~km above sea level for a shower with a zenith angle of 55~degrees observed at: a) the specular angle of the shower axis, the CTR-GM (striped blue line) is seen before the in-air emission (dotted red line), b) 1.7~degrees from the specular angle, the CTR-GM emission is strong and arrives approximately at the same time as the in-air emission, c) 3.9~degrees from the shower axis, the CTR-GM emission is seen after the in-air emission. The situation for a 70 degrees inclined shower at 1.7~degree from the specular angle, is given in d), where it is observed that the CTR-GM emission is negligible. }
  \figlab{results}
\end{figure}

We further note that, being derived from the geomagnetic current, the CTR-GM signal is aligned with the local geomagnetic angle, and thus it has the same polarization--but inverted polarity--to the in-air signal from the induced geomagnetic current, a fact which has so far not been explained without significant tension with the standard model. Additionally, an interesting signature to distinguish the expected CTR-GM signal from the in-air emission is found in the received spectrum, increasing the high-frequency content due to the sharp shock in the potential. As such, a detailed comparison of the spectra of events that satisfy the CTR-GM criteria--anomalous or not--to the spectra of those events which are not expected to be influenced by CTR-GM can be used to confirm the presented hypothesis. 
Finally, since the CTR-GM signal is expected to dominate at small zenith angles we suggest an increased exposure for the ANITA detector towards these angles in future flights which will increase their  cosmic-ray  statistics  within  the  region  of  interest for the CTR-GM signal.

{\bf Conclusions---}
We have shown that coherent transition radiation is not limited to a net excess charge, but applies equally well to a (net charge-neutral) transverse current traversing different media. We investigated if the anomalous events observed by the ANITA detector can be explained by coherent transition radiation from the geomagnetically-induced air shower current hitting the Antarctic surface. It is shown that the anomalous events have a particularly high particle content at the air-ice boundary compared to the typical cosmic-ray events detected by ANITA. Furthermore, we show that for showers with relatively small zenith angles~$\lesssim$~$ 70$~degrees, similar to the two anomalous events detected by ANITA, the expected electric field at the detector can be dominated by coherent transition radiation, and even more interestingly, the apparent polarity can be inverted. For larger zenith angles, the particle content at the boundary is too small to have a significant influence on the expected electric fields. It follows that coherent transition radiation from the geomagnetically-induced air shower current provides a natural, standard model explanation for the ANITA anomalous events. We recommend a more detailed event-by-event investigation to confirm this hypothesis.

{\bf Acknowledgements---} We express our gratitude to the ANITA collaboration for their informative comments and assistance regarding this work. We also thank A. Connolly, D. Besson, A. Romero-Wolf, O. Scholten, S. Buitink, and J. Beacom for edits and insights. This work was supported by the Flemish Foundation for Scientific Research FWO (FWO-12L3715N - K.D. de Vries) and the European Research Council under the EU-ropean Union’s Horizon 2020 research and innovation programme (grant agreement No 805486).\\\\

{\bf Appendix A: Particle distribution parameterization---}
The electric field calculations presented in~\figref{results} are based on the obtained field equations presented in Eq.~6. The three dimensional distribution $J^{x}(t')$ describing the geomagnetically induced air shower current is parameterized as,
\beq
 J^x(t',\vec{r},h)=N_e(t') e v_d w(\vec{r},h).
\eeq
The total particle number as function of the emission time $t'$ is obtained directly from the NKG formalism~\cite{nkg1,nkg2}, and outlined in detail in~\cite{CosmicRayRadio}. Following~\cite{Wer12}, the drift velocity $v_d=0.04\;c$ is taken constant throughout the 3D particle distribution in the cascade front. This distribution is assumed to be radially symmetric and subdivided as $w(\vec{r},h)=2\pi w_1(r)w_2(r,h)$. The radial particle distribution $w_1(r)$ is obtained through,
\beq
w_1(r)=\frac{\Gamma(4.5-p)}{\Gamma(p)\Gamma(4.5-2p)}\left(\frac{r}{r_0}\right)^{p-1}\left(\frac{r}{r_0}+1\right)^{p-4.5}
\eeq
Here $p=1.1$, and $r_0=80$~m are fit parameters fixed to their values obtained close to shower maximum from the CX-MC-GEO package~\cite{Wer12}. The longitudinal particle distribution is a function of distance from the shower axis and given by,
\beq
w_2(r,h)=\frac{4h}{(h_1)^2}\exp(-2h/h_1).
\eeq
The radial dependence is hidden in the width parameter $h_1(r)=5(0.02+0.1r)$~m fitted by comparison of this parameter with results obtained by the CX-MC-GEO package, corresponding to an effective width of 10~cm at the shower axis up to 10 meters at a radial distance of 100~m away from the shower axis (see~Fig [10] of~\cite{Wer12}).

\end{document}